# Eliminating the Transverse Instabilities of Kerr Solitons

Charalambos Anastassiou, [1,2] Marin Soljačić,[1,3] Mordechai Segev,[1,2] Detlef Kip, [1,5] Eugenia D. Eugenieva,[4] Demetrios N. Christodoulides,[4] Ziad H. Musslimani[6], and Juan P. Torres[1,7].

[1] Physics Department and the Solid State Institute, Technion, Haifa 32000, Israel.

[2] Electrical Engineering Department, Princeton University, Princeton, NJ 08544, USA

[3] Physics Department, Princeton University, Princeton, NJ 08544, USA

[4] Electrical Engineering and Computer Science Dept, Lehigh University, Bethlehem, PA 18015, USA

[5] Physics Department, Universität Osnabrück, 49069 Osnabrück, Germany.

[6] Mathematics Department, Technion, Haifa 32000, Israel.

[7] Dept. Signal theory and Communications, Polytechnic University of Catalonia, Jordi Girona 1-3, Campus Nord D3, 08034 Barcelona Spain

**Abstract.** We show analytically, numerically, and experimentally that a transversely stable one-dimensional [(1+1)D] bright Kerr soliton can exist in a 3D bulk medium. The transverse instability of the soliton is completely eliminated if it is made sufficiently incoherent along the transverse dimension. We derive a criterion for the threshold of transverse instability that links the nonlinearity to the largest transverse correlation distance for which the 1D soliton is stable.

Research on optical spatial solitons has made much progress during the past decade: many new systems that support solitons have been identified, solitons of more than one transverse dimension have been demonstrated, and a whole range of soliton interactions was explored [1]. Despite the diversity of the physical systems that support them, solitons are a universal phenomenon and share many common features [1], one of which is Transverse Instability (TI) [2-10]. TI is a symmetry breaking instability: almost all solitons [11] of a particular dimension that propagate in a higher dimension system (by having a uniform wavefunction in the additional dimension/s) are unstable to perturbations in the dimension/s in which they are uniform. TI occurs because perturbations in the dimension of uniformity have nothing to restrain them from growing (driven by the nonlinearity) and breaking the soliton up.

In the particular case of a spatial optical (1+1)D soliton that is self-trapped in one dimension, $x$, is uniform in the transverse dimension, $y$, and is propagating along $z$, TI causes the soliton to break up along $y$ into an array of 2D filaments [2-10]. The transverse-wavelength of these perturbations is usually much larger than the soliton width [2-4]. Transverse instability is especially severe for Kerr nonlinearities, and prohibits spatial 1D Kerr solitons in a bulk medium. This is why spatial (1+1)D Kerr solitons have to be launched in a planar waveguide configuration, in which the $y$-confinement is much narrower than the self-trapped (soliton) width in $x$ [5,6]. TI actually occurs for solitons in any nonlinearity, including, for example, quadratic solitons [7] and photorefractive solitons [8,9]. Interestingly, saturation arrests transverse instability [10] but never completely eliminates it. In fact, it is the suppression of TI due to saturation that facilitates the observation of stable 1D solitons in a bulk photorefractive crystal for more than 10 diffraction lengths [8].

Thus far, in order to avoid TI, experiments with 1D solitons were either conducted in planar waveguides [5,6] or in nonlinearities in which TI was greatly suppressed [8]. Here we demonstrate how to produce a truly stable stripe Kerr soliton propagating in a 3D bulk medium without suffering from transverse instability. We show that if the soliton is made "sufficiently" incoherent in its transverse dimension, $y$, then TI is completely eliminated.

Before proceeding to TI, it is essential to recall incoherent solitons. These are solitons made of partially incoherent light, or self-trapped weakly-correlated wave-packets [12]. They are multimode (speckled) beams of which the instantaneous intensity distribution is varying randomly with time. If such beams are launched into a non-instantaneous self-focusing medium, such that the response time of the nonlinearity greatly exceeds the characteristic fluctuation time, then self-focusing is driven solely by the average intensity. In this manner, the incoherent beam induces a multimode waveguide and traps (guides) itself in it by properly populating the guided modes, thus forming an incoherent soliton [12-18].

A clue that TI could be completely eliminated for solitons was given by two recent discoveries: Modulation Instability (MI) of incoherent light [19] and elliptical incoherent solitons [14,20]. MI belongs to the same family of symmetry breaking instabilities as TI does, and it occurs when a plane wave (or a very broad beam or pulse) is launched into a self-focusing medium. It is well established that if the plane wave is fully coherent, it breaks up into a train of filaments due to MI. Recently, it has been shown theoretically and experimentally [20], that MI does exists also for incoherent light, but it occurs only if the nonlinearity exceeds a well-defined threshold. The MI threshold is determined by the coherence of the light. If the nonlinearity is below threshold, then MI is eliminated and the wave is stable. This generic idea has enabled the observation of anti-dark solitons [21], which were thought to be unstable in conservative nonlinear systems [22]. The new finding of incoherent elliptical solitons, which can exist in logarithmic [14] and saturable [20] nonlinearities, is based on having different coherence properties for the two transverse dimensions of self-trapping (anisotropic coherence function [14,20]). Combing these ideas, one can generate a 1D soliton that is fully coherent in $x$ (direction of trapping), partially incoherent but uniform in $y$, and propagating along z. The intimate relation between TI and MI suggests that TI of incoherent beams should also exhibit a threshold for its existence. Therefore, if the degree of coherence in $y$ is such that TI is below the threshold, all transverse perturbations are suppressed and TI is eliminated altogether. This is core idea of our Letter. The idea of using the threshold to eliminate TI applies to any type of nonlinearity, yet for the rest of this Letter we concentrate on the Kerr nonlinearity for two main reasons. First, wave propagation in Kerr media is described by the cubic nonlinear Schrodinger equation (NLSE) which is one of the most general soliton equations [23]: it describes solitons (at least in some limit) in capillary waves, gravity waves, deep water waves, shallow water waves, plasma waves, and optics. Generally speaking, the NLSE describes envelope solitons in dispersive wave systems with weak symmetric anharmonicity. The second reason is that the effect of TI for Kerr solitons is very strong and we can demonstrate a convincing difference between having TI and eliminating TI by making the soliton incoherent along $y$.

To formalize the problem, we use the radiation transfer method [17]. An incoherent beam can be considered as a series of coherent speckles that change, on average, every coherence time, $\tau_{coh}$. We define $B(x_1,y_1,x_2,y_2,z) = <E^*(x_2,y_2,z,t)E(x_1,y_1,z,t)>$, the spatial correlation function, where $E(x,y,z,t)$ is the slowly varying amplitude. The brackets denote averaging over the response time of the medium, $\tau$, which is much larger than $\tau_{coh}$. From the paraxial wave equation we get [17]

$$\frac{\partial B}{\partial z} - \frac{i}{k}\{\frac{\partial^2 B}{\partial x \partial \rho_x} + \frac{\partial^2 B}{\partial y \partial \rho_y}\} = \frac{ik}{n_0}\{\delta n(x_1,y_1,z) - \delta n(x_2,y_2,z)\}B, \qquad (1)$$

where $z$ is the propagation direction, $k$ the carrier wavenumber, $n_0$ the bias refractive index, $\delta n$ the nonlinear contribution to the refractive index, $x = (x_1+x_2)/2$, $y = (y_1+y_2)/2$ the middle point coordinates, and $\rho_x=x_1-x_2$, $\rho_y=y_1-y_2$ the difference coordinates. When $\rho_x=\rho_y=0$, $B(x,y,\rho_x,\rho_y,z)$ is the time-averaged light intensity, $I(x,y,z)$. Let $B_S(x,y,\rho_x,\rho_y)$ = $u(x+\rho_x/2)u^*(x-\rho_x/2)A_n(\rho_y)$ be a stationary ($z$-independent) solution of Eq. (1). It represents a 1D soliton stripe, which is self trapped and fully coherent in $x$, while being uniform and incoherent along $y$ with an angular spectrum of $A_n(\rho_y)$. The waveform $u(x)$ is determined by the nonlinearity, and can be taken to be real without loss of generality. In Kerr media, $u(x)$ is proportional to $sech(x)$. To study TI, we add a small perturbation $B_I$ to $B_S$ where $B_I<<B_S$. The nonlinear index change in Kerr media is $\delta n(I) = \gamma I$, where $\gamma$ is the nonlinear coefficient ($n_2$). Linearizing Eq. (1) yields

$$\frac{\partial B_1}{\partial z} - \frac{i}{k}\{\frac{\partial^2 B_1}{\partial x \partial \rho_x} + \frac{\partial^2 B_1}{\partial y \partial \rho_y}\} =$$

$$\frac{ik}{n_0}\gamma\left\{\begin{array}{l}[B_1(x+\frac{\rho_x}{2},y+\frac{\rho_y}{2},\rho_x=0,\rho_y=0)-B_1(x-\frac{\rho_x}{2},y-\frac{\rho_y}{2},\rho_x=0,\rho_y=0,z)]B_S(x,y,\rho_x,\rho_y)+\\ [B_S(x+\frac{\rho_x}{2},y+\frac{\rho_y}{2},\rho_x=0,\rho_y=0)-B_S(x-\frac{\rho_x}{2},y-\frac{\rho_y}{2},\rho_x=0,\rho_y=0),z]B_1(x,y,\rho_x,\rho_y)\end{array}\right\}$$

(2)

We seek solutions in the form $B_1(x,y,\rho_x,\rho_y,z)=exp(gz)exp(i\alpha y)L(x,\rho_x)A_f(\rho_y) + exp(g^*z)exp(-i\alpha y) L^*(x,-\rho_x)A_f^*(-\rho_y)$, where $\alpha$ is the transverse wavenumber, $g$ the TI growth rate (gain), and $A_f(\rho_y)$ the angular spectrum. The necessary condition $B_1(x,y,\rho_x,\rho_y,z)= B_1^*(x,y,-\rho_x,-\rho_y,z)$ is satisfied [17,19]. Substituting $B_1$ into Eq. (2) gives:

$$gL(x,\rho_x)A_f(\rho_y) - \frac{i}{k}\{\frac{\partial^2 L(x,\rho_x)}{\partial x \partial \rho_x}A_f(\rho_y) + i\alpha\frac{dA_f(\rho_y)}{d\rho_y}L(x,\rho_x)\} =$$

$$\frac{ik}{n_0}\gamma\left\{\begin{array}{l}[L(x+\frac{\rho_x}{2},0)A_f(0)e^{i\alpha\rho_y/2} - L(x-\frac{\rho_x}{2},0)A_f(0)e^{-i\alpha\rho_y/2}]u(x+\frac{\rho_x}{2})u(x-\frac{\rho_x}{2})A_n(\rho_y)\\ + [u(x+\rho_x/2)^2 - u(x-\rho_x/2)^2]L(x,\rho_x)A_f(\rho_y)A_n(0)\end{array}\right\}$$

(3)

We are interested in determining the threshold condition, that is, to find the conditions under which the growth rate $g(\alpha)$ goes from a positive value to a negative value for all $\alpha$. The procedure of determining the threshold applies to any form of spatial coherence (angular power spectrum), but for simplicity, we consider an initial Gaussian angular power spectrum, i.e. $A_n(\rho_y) = exp[-(\rho_y\theta_0 k/2)^2]$, where $\theta_o$ defines the degree of coherence (the correlation distance). The higher $\theta_o$ the more incoherent the soliton is. For a fully coherent soliton, if we were to calculate the growth rate $g$ as a function of transverse wavenumber $\alpha$, then $g$ starts from 0 (at $\alpha=0$), increases and reach a maximum positive value (at the wavenumber with the largest growth rate), and then drops back to 0 at $\alpha$ associated with the "cutoff wavelength" [2-4]. This means that for a coherent soliton, the growth rate is positive (and TI exists) for a band of wavenumbers $\alpha$ between zero and the cutoff wavenumber. For a soliton that is partially-coherent in $y$, we expect that if TI occurs, then $g(\alpha)$ will be positive for a band of wavenumbers, just like in the coherent case. On the other hand, if TI does not occur, then $g(\alpha)$ will be negative for all $\alpha$. In other words, if we start from $\theta_o$ small enough (a beam that is coherent enough), $g(\alpha)$ will be positive in a band of wavenumbers, just like the coherent case. But, as $\theta_o$ increases, this band becomes narrower until it completely disappears at some value $\theta_{oT}$. If $\theta_o$ is larger than this value, then transverse instability is completely eliminated. [Recall that modulation instability of incoherent light exhibits a similar trend: see Fig. 1 in [19], which shows $g(\alpha)$ of the incoherent MI as the threshold is approached]. We therefore expect that, at the threshold $\theta_o = \theta_{oT}$, the two boundary points at which $g(\alpha)=0$ (one at $\alpha=0$ and the other at the cutoff wavenumber), coincide. Thus, we seek the value of $\theta_o$ at which (I) $g(\alpha=0)=0$ and (II) $g'(\alpha=0) = 0$. We solve Eq. (3) by expansion while retaining up to $O(\alpha)$. This eliminates the first term in the LHS because $g = O(\alpha^2)$. For Kerr solitons, the growth rate of the transverse instabilities, $g$, is independent of $x$ [2-4], that is, the growth rate of the transverse perturbations is the same across the 1D soliton (even though the actual shape of the perturbations depends on $x$). We can therefore seek solutions of Eq. 3 under the conditions (I) and (II) at the center of the soliton, i.e., at $x = \rho_x = 0$, and safely assume that the threshold we find is the same everywhere on the soliton. It can be easily shown (by expanding into derivatives with respect to $x_1$ and $x_2$) that the second term of the LHS is zero for $x_1 = x_2$ i.e. for $\rho_x = 0$. Thus, from Eq. (3) we get:

$$\frac{\alpha}{k}\frac{dA_f(\rho_y)}{d\rho_y}L(0,0) = -\frac{kI_0}{n_0}\gamma L(0,0)\alpha\rho_y A_f(0)\exp[-(k\theta_{0T}\rho_y/2)^2] \qquad (4)$$

where $I_0 = u(0)^2$ is the peak intensity of the soliton. On physical grounds it is unlikely that a small perturbation will alter the coherence statistics of the soliton (especially here that propagation effects, given by $g$, are of order of $\alpha^2$ and are ignored). Thus, we can safely assume that, $A_f(\rho_y) = A_n(\rho_y)$. Eq. (4) gives the threshold condition $\theta_{0T}^2 = 2\Delta n_0/n_0$, where $\Delta n_0 = \gamma I_0$ is the maximum index of refraction change. One can actually calculate, using numerical methods similar to those of [10], the entire function $g(\alpha)$ and from it obtain the threshold for any angular distribution function [25].

To verify the analytic predictions, we perform numerical simulations using the coherent density approach [12]. We launch a 1D Kerr soliton with a Gaussian angular power spectrum, $A(\rho_y) = \exp[-(\rho_y \theta_0 k/2)^2]$, for various values of $\theta_0$. In this example, $n_0 = 2.3$, $\lambda$ in vacuum is 0.5 μm, FWHM = 9 μm, which yields a *Δn₀ = 0.0001056* and an analytic prediction of the threshold of $\theta_{0T}$ = 0.55°. Our numerical results are displayed in Fig. 1, where we show images of the intensity distribution of the soliton and cross sections of the intensity along *y* for *x* = 0. In Fig. 1(a) we show the input soliton at *z* = 0. In 1(b) we show a fully coherent soliton for ($\theta_0$ = 0°) after 0.8 cm of propagation. As clearly depicted in the figure, TI breaks the soliton up into a train of 2D filaments. As we approach the threshold, the TI gain is getting smaller: as we set $\theta_0$ to 0.5°, it takes a 4.5cm propagation to exhibit signs of TI [Fig. 1(c)]. To show that TI is completely eliminated when the nonlinearity is below threshold, we increase $\theta_0$ to 0.56°. As shown in 1(d), after 4.5 cm of propagation there are absolutely no signs of TI.

Our experiments are conducted in photorefractive SBN:75 crystal in a setup similar to that of [12]. The beam is made spatially incoherent by passing it through a rotating diffuser. The rotating diffuser provides a new phase and amplitude distribution every $\tau_{coh}$ ~ 1 μs, which is much shorter than the response time of the medium $\tau$ ~ 1 s. Unlike all previous experiments with incoherent solitons, here we need to generate a beam which is very narrow and fully coherent in *x*, yet uniform and partially incoherent *y*. To do that, we use a cylindrical lens which focuses the beam only in the *y* direction onto the rotating diffuser. Then, by moving the focal point of this lens closer (farther away) from the diffuser, we increase (decrease) the coherence in *y*. The *x* coherence is not affected by the translation of this lens. After the diffuser, the beam is collimated (to ~ 2 cm) and passed through a narrow (along *x*) slit. The slit is made narrower than the speckle size in *x*, and it effectively creates a 1D beam that is narrow and coherent in *x* and "infinitely" long (uniform) and incoherent in *y*. The slit is then imaged to the input face of the crystal. We get a reasonable estimate of the correlation distance by stopping the diffuser and measuring the average speckle size at the crystal input plane. Finally, we use an orthogonally polarized background beam that covers the crystal uniformly as necessary for photorefractive screening solitons [8]. The input and output faces of the crystal are imaged onto a CCD camera.

The photorefractive screening nonlinearity is in general saturable but resembles the Kerr nonlinearity when the peak intensity of the soliton normalized to the background intensity (or the dark irradiance) is much smaller than unity [8,24]. In our case, this ratio is ~ 0.1. At this normalized intensity, a soliton that is fully coherent in both $x$ and $y$ exhibits very strong transverse instability, as shown experimentally in [8]. We can therefore test the main prediction of our theory by gradually increasing the incoherence in $y$ (decreasing the speckle size) until the soliton becomes transversely stable. All other parameters (applied field, intensity) are kept constant. Our results are shown in Fig. 2. The 12 μm FWHM input beam [Fig. 2(a)] linearly-diffracts to a 60 μm FWHM output beam after 6 mm of propagation in the crystal [Fig. 2(b)]. The nonlinearity is turned on with the application of 2.7 kV/cm and the beam self-traps forming a soliton in $x$. When the beam is fully coherent, then the soliton suffers from TI and breaks up into filaments [Fig. 2(c)]. When the beam is made incoherent in $y$, but with a large speckle size (~100 μm), then the nonlinearity is still above the TI threshold and beam still suffers from TI [Fig. 2(d)]. Finally, by decreasing the speckle sizes to ~ 5 μm, TI is eliminated and we get a stable (1+1)D soliton [Fig. 2(e)]. Thus, we have shown that a stable (1+1)D soliton can propagate in a 3D bulk medium provided that it is made partially-incoherent in the "uniform" transverse dimension. For the soliton to be stable, the degree of coherence in the "dimension of uniformity" must be such that the nonlinearity is below the threshold for transverse instability.

In conclusion, we have derived theoretically the threshold for TI of (1+1)D solitons that are fully coherent in their direction of trapping yet are partially-incoherent in their direction of uniformity. We predicted that if the nonlinearity is below a well-defined threshold, then transverse instability of such 1D solitons is completely eliminated. We proved our results analytically, numerically, and experimentally, and specifically showed that it is possible to generate stable 1D Kerr-like solitons in a 3D bulk material. This is the only method we know of for propagating truly stable 1D solitons in a bulk material. We note that in principle our method applies to all types of saturable nonlinearities, and could be used to eliminate TI in them as well. Finally, we believe that this work opens up a whole range of possibilities of eliminating transverse instabilities in many soliton systems, for example, instabilities of 1D dark solitons in bulk media, instabilities of ring beams (with and without topological charge) in self-focusing media, and more.

This work was supported by the Israeli Science foundation, the US Army Research Office, the NSF, and AFOSR. It is part of the MURI program on optical spatial solitons. E. Eugenieva acknowledges a NSF-NATO fellowship in Science and Engineering. This research was supported in part by the Pittsburgh Supercomputing Center, Pittsburgh, PA.

**FIGURE CAPTIONS:**

**Figure 1:** Numerical simulations of a 1D Kerr soliton with a Gaussian angular spectrum, for various degrees of coherence. Here, $\lambda = 0.5$ μm, $n_0=2.3$, the beam FWHM is 9 μm, and $\Delta n_0 = 0.0001056$ The analytically-predicted threshold is $\theta_0=0.549°$. (a) The input intensity. (b) The output beam after 0.8 cm of propagation for a fully coherent beam $\theta_0 = 0°$, showing that the soliton is destroyed by TI. (c) The output beam close to the threshold (for $\theta_0 = 0.5°$), after 4.5 cm of propagation: as the threshold is approached, the TI gain is reduced and it takes a longer propagation distance before TI is evident. (d) The output beam for $\theta_0 = 0.56°$, which is below the threshold, after 4.5 cm of propagation. **TI is completely eliminated**.

**Figure 2:** Experimental results in a photorefractive SBN:75 crystal in the Kerr regime (intensity ratio ~ 0.1). (a) The input 12 μm FWHM beam. (b) The output beam after 6 mm of linear diffraction when nonlinearity is off. (c), (d), (e) show the output beams with an applied field of 2.7 kV/cm, for various degree of coherence along $y$ and all other parameters kept constant. (c) A fully-coherent soliton breaks up into filaments because of TI. (d) The soliton is made incoherent along $y$ but with large speckle sizes ~ 100 μm (small $\theta_0$) and still displays a strong TI. (e) The speckle sizes are reduced to ~ 10 μm: **TI is completely eliminated**.

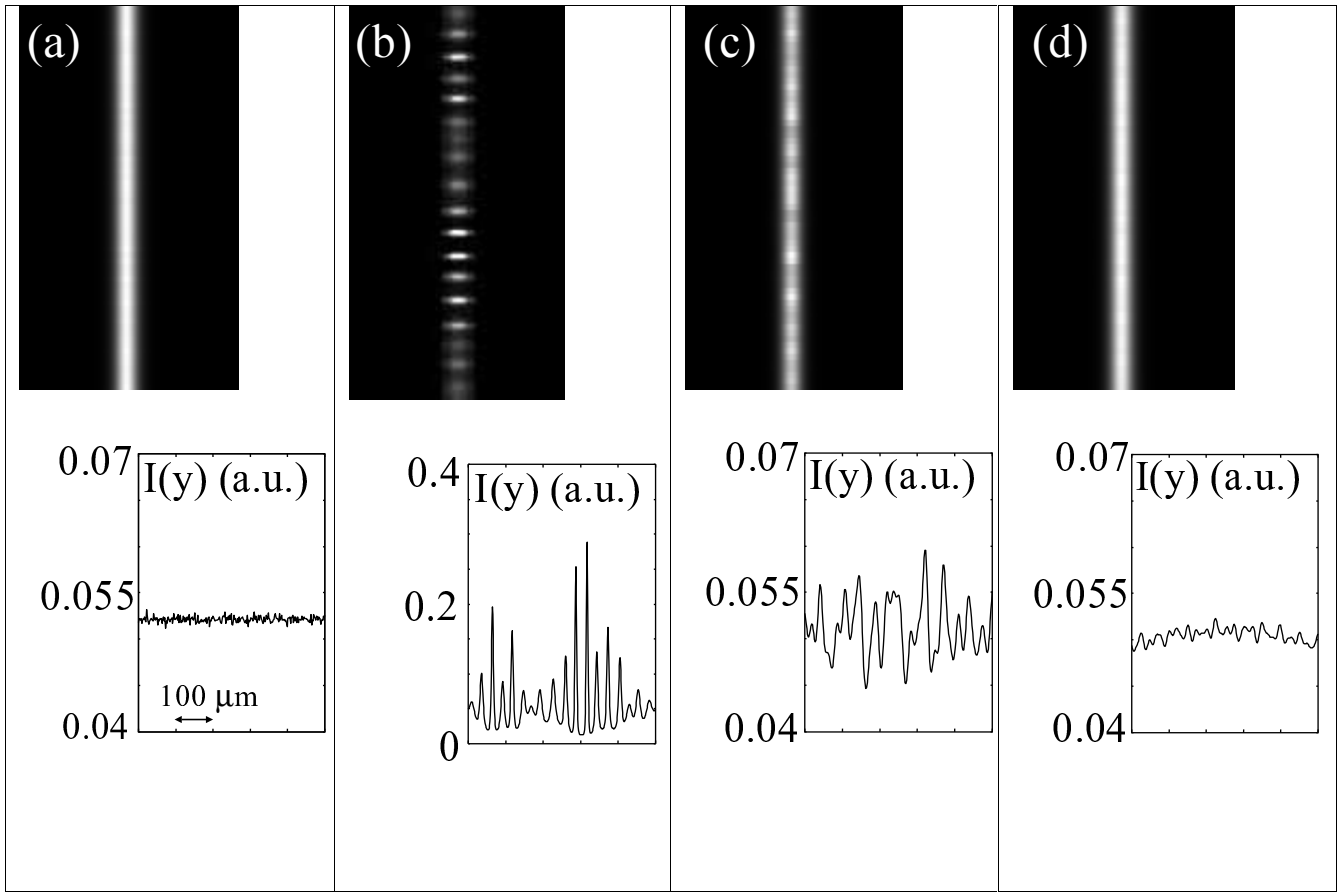

Fig. 1

Charalambos Anastassiou et.al.

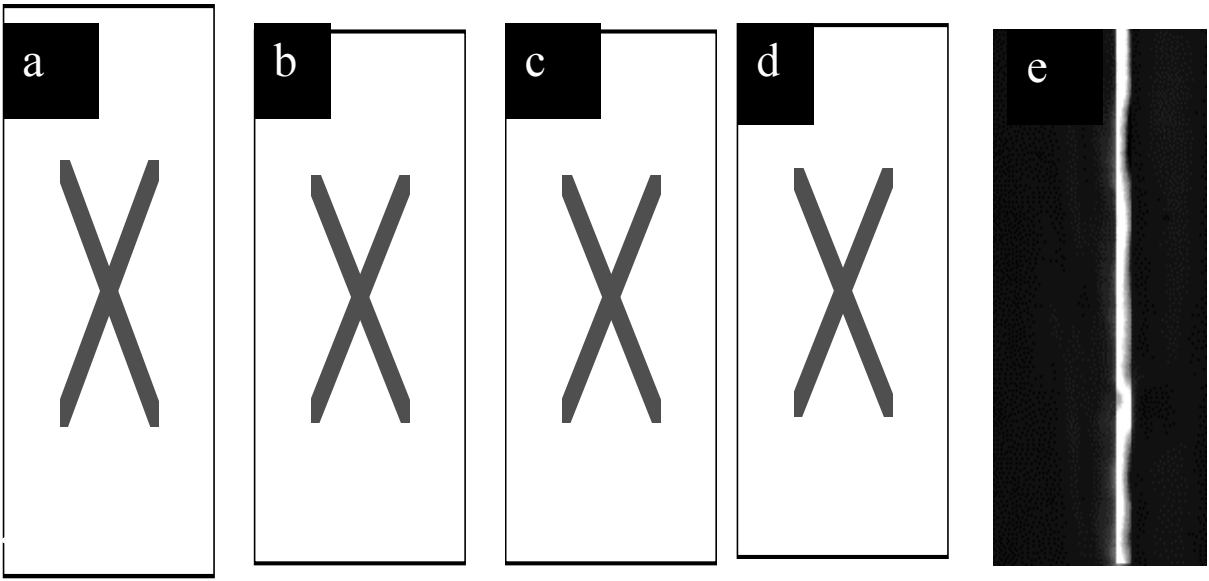

Fig. 2

Charalambos Anastassiou et. al.